\documentclass[12pt]{article}

\def\lsim{\mathrel{\rlap{\lower3pt\hbox{\hskip0pt$\sim$}}
     \raise1pt\hbox{$<$}}}         
\def\gsim{\mathrel{\rlap{\lower4pt\hbox{\hskip1pt$\sim$}}
     \raise1pt\hbox{$>$}}}         

\usepackage{amsmath}
\usepackage{amsfonts}

\begin{document}
\begin{titlepage}

\centerline{\Large \bf Massive Gravity in Extra Dimensions}
\medskip

\centerline{Zurab Kakushadze$^\S$$^\dag$$^\P$\footnote{\tt Email: zura@quantigic.com}}

\bigskip

\centerline{\em $^\S$ Quantigic$^\circledR$ Solutions LLC}
\centerline{\em 1127 High Ridge Road \#135, Stamford, CT 06905\,\,\footnote{DISCLAIMER: This address is used by the corresponding author for no
purpose other than to indicate his professional affiliation as is customary in
scientific publications. In particular, the contents of this paper are limited
to Theoretical Physics, have no commercial or other such value,
are not intended as an investment, legal, tax or any other such advice,
and in no way represent views of Quantigic$^\circledR$ Solutions LLC,
the website {\underline{www.quantigic.com}} or any of their other affiliates.}}
\centerline{\em $^\dag$ Department of Physics, University of Connecticut}
\centerline{\em 1 University Place, Stamford, CT 06901}
\centerline{\em $^\P$ Department of Theoretical Physics, A. Razmadze Mathematical Institute}
\centerline{\em I. Javakhishvili Tbilisi State University}
\centerline{\em 6 Tamarashvili Street, 0177 Tbilisi, Georgia}
\medskip
\centerline{(March 10, 2014; revised May 13, 2014)}

\bigskip
\medskip
\centerline{\it Dedicated to the memory of my teacher and mentor}
\centerline{\it Prof. Revaz Dogonadze (Nov 21, 1931 -- May 13, 1985)}

\bigskip
\medskip

\begin{abstract}
{}We discuss a Brane World scenario where we live on a 3-brane with massive gravity in infinite-volume bulk. The bulk graviton can be much heavier than the inverse Hubble size, as heavy as the bulk Planck scale, whose lower bound is roughly the inverse of 0.1 mm. The 4D Einstein-Hilbert term on the brane shields the brane matter from both strong bulk gravity and large bulk graviton mass. Gravity on the brane does not become higher-dimensional at large distances. Instead, at distance scales above the bulk Planck length, gravity on the brane behaves as 4D gravity with small graviton mass roughly of order or below the inverse Hubble size. Unlike the massless case, with massive gravity in the bulk one can have: i) 4D tensor structure on a codimension-1 brane; and ii) no infrared tachyon for smoothed-out higher codimension branes. The effects of the brane dynamics on the bulk are exponentially suppressed away from the brane. One consequence is that there are no ``self-accelerated" solutions. In codimension-2 cases there exist nonsingular solutions with a flat 3-brane for a continuous positive range of the brane tension. In higher codimension cases, as in the massless case, higher curvature terms are required to obtain such solutions.
\end{abstract}
\medskip
PACS: 04.50.-h, 04.50.Gh, 04.50.Kd, 11.25.-w, 11.27.+d
\end{titlepage}

\newpage

\section{Introduction and Summary}

{}One motivation for massive gravity is the currently observed accelerated expansion of the Universe \cite{Nova, Nova1}. Small graviton mass might lead to a large-scale modification of gravity at Hubble distance scales with an accelerated expansion sans a tiny cosmological constant. Another motivation is string theory description of QCD, where massless spin-2 modes present in all known consistent string theories would somehow have to acquire mass \cite{thooft}.

{}In this note we propose another application of massive gravity. In the Brane World scenario we live on a 3-brane embedded in a higher-dimensional bulk. Infinite-volume bulk is particularly interesting for a multitude of reasons, including the cosmological constant problem \cite{DGP, Witten, DG}. One aims at infinite-volume solutions with a non-inflating (or very slowly inflating) brane and the property that such solutions exist for a continuous range
of positive 3-brane tension.\footnote{\, Such solutions do not exist for codimension-1 branes \cite{DGP}.} One difficulty is that, when bulk gravity is massless, the tensor structure of the graviton propagator on a codimension-1 brane is 5-dimensional \cite{DGP}. The correct 4-dimensional tensor structure can be obtained on codimension-2 and higher branes \cite{DG}. In these cases singularities in the graviton propagator, even if the brane tension vanishes so the background is nonsingular, must be smoothed out \cite{Orientiworld, DGHS}, which introduces ``unconventional" states, {\em e.g.}, a tachyon, albeit with small negative mass-squared or order (sub-)inverse-Hubble-size-squared \cite{Orientiworld}.\footnote{\,Such a tachyon may not necessarily be problematic, and may further be an artifact of neglecting an infinite series of non-local terms on the brane, which are expected to arise once we smooth out the brane and be suppressed by powers of the brane ``thickness" \cite{Orientiworld}.} Solutions with flat brane exist for a continuous range of positive brane tension in the codimension-2 case; however, precisely in the codimension-2 case there is an upper limit on the brane tension such that it does not improve the experimental bound on the 4-dimensional cosmological constant \cite{6D2}. This is circumvented in higher-codimension cases, which, however, require inclusion of higher-curvature terms in the bulk to avoid naked singularities \cite{CIK}.

{}Our proposal is to have {\em massive} gravity in the bulk. The bulk graviton can be much heavier than the inverse Hubble size, as heavy as the bulk Planck scale, whose lower bound is roughly the inverse of 0.1 mm. The 4D Einstein-Hilbert term on the brane shields the brane matter from both strong bulk gravity and large bulk graviton mass. Gravity on the brane does not become higher-dimensional at large distances. Instead, at distance scales above the bulk Planck length, gravity on the brane behaves as 4D gravity with small graviton mass roughly of order or below the inverse Hubble size. Unlike the massless case, with massive gravity in the bulk one can have: i) 4D tensor structure on a codimension-1 brane; and ii) no infrared tachyon for smoothed-out higher codimension branes.

{}Our basic proposal can be summarized by the following action:
\begin{equation}\label{BW}
 S = {\widehat M}_P^2 \int_\Sigma d^4 x \sqrt{-{\widehat G}}\left[{\widehat R} - {\widehat f}+ \dots\right] + M_P^{D-2} \int d^D x \sqrt{-G}\left[R - \mu^2 V\right]~,
\end{equation}
where ${\widehat M}_P$ is the 4-dimensional Planck scale, $f\equiv {\widehat M}_P^2 {\widehat f}$ is the 3-brane tension, ${\widehat G}_{\mu\nu} \equiv {\delta_\mu}^M {\delta_\nu}^N G_{MN}|_\Sigma$ ($x^\mu$, $\mu = 0,1,2,3$ are the coordinates in the brane worldvolume $\Sigma$), the brane scalar curvature ${\widehat R}$ is constructed from ${\widehat G}_{\mu\nu}$, the ellipses stand for brane matter terms, $M_P$ is the bulk Planck scale, $\mu$ is a mass parameter,\footnote{\, Not to be confused with a worldvolume index. Typically, the bulk graviton mass $M\sim \mu$.} and $V$ is a dimensionless ``potential" for the metric $G_{MN}$ that makes bulk gravity massive ($x^M$, $M=0,1,\dots,(D-1)$ are the coordinates in the bulk, and the bulk scalar curvature $R$ is constructed from $G_{MN}$; $M$ not to be confused with the bulk graviton mass). {\em {A priori}} $D$ is arbitrary, albeit in this paper we will mostly discuss codimension-2 solutions (both point- and string-like 3-branes in 6D bulk), and also other cases.

{}Conceptually, for our proposal here it is not critical what the ``potential" $V$ is or how it arises. For the sake of concreteness, however, here we assume that $V$ arises via the gravitational Higgs mechanism, where graviton acquires mass via spontaneous breaking of diffeomorphisms by scalar vacuum expectation values.\footnote{\, Here we deliberately will not delve into the ongoing debate in the literature on consistency of massive gravity for this is outside of the scope of this note. The author's current view on this subject is summarized in \cite{ZK.massive}, which also contains many relevant references. For a general review of the Brane World scenario in the string theory context, see, {\em e.g.}, \cite{KT} and references therein.}

{}The simplest model of the gravitational Higgs mechanism is that of \cite{thooft}. In this model, upon gauging away the scalars, the ``potential" $V$ is given by \cite{Unitarity, ZK.massive}:\footnote{\, In this model the naive perturbative mass term is not of the Fierz-Pauli form. Nonetheless, non-perturbatively the theory ``resums'' and the Hamiltonian is positive-definite \cite{JK, Unitarity}.}
\begin{equation}\label{tHIK}
 V = G^{MN}\eta_{MN} - (D-2)~,
\end{equation}
where $\eta_{MN}$ is the Minkowski metric. We use this model to construct explicit codimension-2 Brane World solutions of (\ref{BW}), where a flat (Minkowski) 3-brane exist for a continuous range of the brane tension $0\leq f < f_c$, where $f_c \equiv 4\pi M_P^4$. The bulk graviton mass can be as large as $M_P$. We also discuss the graviton propagator on the brane in various codimensions for other potentials $V$, including with the Fierz-Pauli perturbative mass term, in which case perturbative expansion actually breaks down.

\section{Non-perturbative Static Solutions}

{}In the next three sections we focus on the model (\ref{tHIK}):
\begin{equation}\label{MG.action}
 S_{MG} = M_P^{D-2}\int d^Dx \sqrt{-G}\left[ R + \mu^2\left(D - 2 - G^{MN}{\widetilde G}_{MN}\right)\right]~,
\end{equation}
where ${\widetilde G}_{MN}$ is the background metric. The equations of motion read
\begin{equation}\label{R.eom}
 R_{MN} = \mu^2\left[ {\widetilde G}_{MN} - G_{MN} \right]
\end{equation}
with the Bianchi identity
\begin{equation}\label{phi.eom}
 \partial_M\left[\sqrt{-G} G^{MN}{\widetilde G}_{NS}\right] - {1\over 2}\sqrt{-G} G^{MN}\partial_S {\widetilde G}_{MN} = 0~,
\end{equation}
which is equivalent to the gauge-fixed equations of motion for the scalars. In (\ref{MG.action}) we have deliberately omitted any source terms. The background metric ${\widetilde G}_{MN}$ is assumed to be Ricci-flat (so $G_{MN} \equiv {\widetilde G}_{MN}$ solves (\ref{R.eom})) away from the sources.

{}Despite its apparent simplicity, the ``mass term" ({\em i.e.}, the r.h.s.) in (\ref{R.eom}) makes it even more nontrivial to solve exactly. Exact massive time-dependent vacuum solutions (mostly in $D=3$ and some for all $D$) were recently constructed in \cite{ZK.massive}. In this note we discuss non-perturbative static solutions, first in $D=3$ for point-like sources, and then codimension-2 Brane World solutions for 3-branes in $D=6$.

\subsection{Massless Case}\label{sub.massless}

{}Let us set up our framework by considering the massless case in $D=3$. The equations of motion
\begin{equation}
 R_{MN} = 0
\end{equation}
admit the following solution:
\begin{equation}\label{metric.1}
 ds^2 = - dt^2 + \exp(2\omega)~\delta_{ij}dx^i dx^j~,
\end{equation}
where $x^i$, $i=1,2$ are the spatial coordinates, $\omega$ is independent of $t$ and {\em away from the origin} (see below) satisfies the following equation:
\begin{equation}
 \partial^i\partial_i\omega = 0~.
\end{equation}
A non-trivial solution is given by:
\begin{equation}
 \omega = -{1\over 8\pi} {\widetilde f}~\ln\left(x^2\over a^2\right)~.
\end{equation}
Here $x^2\equiv x^i x_i$ (the spatial indices are raised and lowered with $\delta^{ij}$ and $\delta_{ij}$, respectively) and $a$ is an integration constant. The meaning of ${\widetilde f}$ becomes evident from the observation that $\omega$ satisfies the following equation, including at the origin:
\begin{equation}
 \partial^i\partial_i\omega = -{1\over 2}~{\widetilde f}~\delta^{(2)}(x^i)~.
\end{equation}
This implies that there is a point-like source at the origin. The corresponding source term in the action reads
\begin{equation}\label{source}
 S_{source} = - f \int_\Sigma d\tau~,
\end{equation}
where the integral over the proper time $d\tau \equiv dt\sqrt{-{\widehat G}_{00}}$ is over the worldline $\Sigma$, ${\widehat G}_{00} \equiv G_{00}|_\Sigma$, and $f\equiv M_P{\widetilde f}$ is the mass of the point-like source.\footnote{\,We use $f$ instead of $m$ to avoid confusion with the graviton mass.}

{}The metric (\ref{metric.1}) is actually flat except at the origin. Let $x^1\equiv \rho\cos(\phi)$, $x^2\equiv\rho\sin(\phi)$ and
\begin{equation}
 r \equiv {1\over {1-\nu}} a^\nu\rho^{1-\nu}~,
\end{equation}
where $\nu\equiv {\widetilde f}/4\pi$ and we are assuming $\nu < 1$. In the $(r,\phi)$ coordinates the metric
\begin{equation}
 ds^2 = -dt^2 + dr^2 + (1-\nu)^2 r^2d\phi^2~.
\end{equation}
In the spatial directions we have a 2-dimensional ``wedge", which is flat away from the origin (and all the curvature is localized at the source),\footnote{\, For $1-\nu = 1/N$, where $N$ is an integer, the ``wedge" is nothing but the $R^2/Z_N$ orbifold with the origin identified with the orbifold fixed point.} with the deficit angle $\theta = 2\pi\nu = {\widetilde f} / 2$. This implies that $0\leq f < f_c$, and the deficit angle $\theta = 2\pi$ for the critical mass $f_c \equiv 4\pi M_P$.

\subsection{Massive Case}

{}Let us now turn to the equations of motion for the massive case (\ref{R.eom}). They still admit a solution of the form (\ref{metric.1}). Away from the origin the equation for $\omega$ reads:
\begin{equation}\label{eq.omega}
 \partial^i\partial_i\omega = \mu^2~\left[\exp(2\omega)- 1\right].
\end{equation}
Here we are interested in $\phi$-independent solutions, which satisfy
\begin{equation}\label{eq.y}
 \partial_\sigma^2 y + {1\over \sigma}\partial_\sigma y = \exp(y) - 1~,
\end{equation}
where $y\equiv 2\omega$, and $\sigma\equiv\sqrt{2}\mu\rho = M\rho$ is a dimensionless variable ($M$ is the perturbative graviton mass). We will analyze this equation in two steps.

\subsection{Naive Linearized Approximation}

{}First, let us consider a naive linearized approximation where one assumes that $y$ is small and keeps only linear terms in $y$:
\begin{equation}\label{eq.z}
 \partial_\sigma^2 z + {1\over \sigma}\partial_\sigma z = z~,
\end{equation}
where we use $z$ instead of $y$ for a solution to the linearized equation. Two independent solutions to (\ref{eq.z}) are given by the modified Bessel functions
$I_0(\sigma)$ and $K_0(\sigma)$, where the former grows exponentially at large $\sigma$ (and consequently must be discarded), and the latter decays exponentially at large $\sigma$ (and therefore is physically relevant). The physical solution is given by
\begin{equation}
 z = {1\over 2\pi}{\widetilde f}~K_0(\sigma)~.
\end{equation}
Noting the small $\sigma$ behavior $K_0(\sigma)\sim -\ln(\sigma/2) - \gamma$ ($\gamma$ is the Euler constant), $\omega\equiv z/2$ satisfies the following equation including at the origin:
\begin{equation}
 \partial^i\partial_i\omega - 2\mu^2\omega = -{1\over 2}~{\widetilde f}~\delta^{(2)}(x^i)~.
\end{equation}
As in Subsection \ref{sub.massless}, this implies that there is a point-like source at the origin with the source term in the action given by (\ref{source}), where again $f = M_P {\widetilde f}$.

{}Near the origin ($\sigma\ll 1$) the space is a 2-dimensional ``wedge" with the deficit angle $\theta = {\widetilde f}/2$ (so we have the critical mass $f_c = 4\pi M_P$, same as in Subsection \ref{sub.massless}). At $\sigma\gg 1$, we have the following leading asymptotic behavior:
\begin{equation}
 \omega \sim {{\widetilde f}\over 4\sqrt{2\pi\sigma}}\exp(-\sigma)~,
\end{equation}
so the space is asymptotically flat with no deficit angle.

\subsection{Non-perturbative Analysis}\label{nonpert}

{}We now turn to the full non-perturbative equation (\ref{eq.y}). We are interested in solutions where $y\rightarrow 0+$ at large $\sigma$. It then follows that $y$ is positive for all $\sigma$. Indeed, if $y$ flips sign at some finite $\sigma_1$, then at some $\sigma_2 > \sigma_1$ we must have $\partial_\sigma y(\sigma_2) = 0$ and $y(\sigma_2) > 0$, which according to (\ref{eq.y}) would imply $\partial^2_\sigma y(\sigma_2) > 0$, which in turn is incompatible with $y < 0$ at $\sigma <\sigma_1$ and $y > 0$ at $\sigma > \sigma_2$. So, the solutions we are looking for are positive at all $\sigma$ and decay to zero at large $\sigma$. Such solutions indeed exist and can be constructed numerically by noting that at large $\sigma$ (\ref{eq.y}) is well-approximated by the linearized equation (\ref{eq.z}), so the values of $y$ and $\partial_\sigma y$ at some $\sigma = \sigma_* \gg 1$ can be taken to be $y(\sigma_*) = ({\widetilde f}_1/2\pi)K_0(\sigma_*)$ and $\partial_\sigma y(\sigma_*) = ({\widetilde f}_1/2\pi)K_0^\prime(\sigma_*)$, and the second-order ordinary differential equation (\ref{eq.y}) can be solved using the standard numeric methods (Euler, Runge-Kutta, {\em etc.}). Here, however, the relation of ${\widetilde f}_1$ to the mass at the origin is not as simple as in the linearized case (see below).

{}While the behavior at large $\sigma$ is well-approximated by the solution to the linearized equation, the near-origin behavior is affected by the exponent on the r.h.s. of (\ref{eq.y}), which grows more rapidly than the naive linearized mass term in (\ref{eq.z}) near the origin, where $y$ goes to infinity. However, precisely because the exponent dominates in the near-origin regime, we can analyze it analytically. Near the origin the relevant solution to (\ref{eq.y}) is well-approximated by a solution to the following equation
\begin{equation}
 \partial_\sigma^2 u + {1\over\sigma}\partial_\sigma u = \exp(u)
\end{equation}
Let $\sigma\equiv \exp(-\zeta)$ and $u\equiv 2(v + \zeta)$. We have:
\begin{equation}
 2\partial_\zeta^2 v = \exp(2v)~.
\end{equation}
The relevant solution is given by
\begin{equation}\label{v}
 v = \ln\left[\sqrt{2}b\over\sinh(b(\zeta - \zeta_*))\right]~,
\end{equation}
where $b>0$ and $\zeta_*$ are integration constants. (For $b = 0$, $v = -\ln[(\zeta - \zeta_*)/\sqrt{2}]$.)

{}\noindent For the $b > 0$ solution the near-origin behavior is as follows:
\begin{equation}
 u \sim -2(1-b)\ln(\sigma)~.
\end{equation}
This implies that at the origin we have a point-like source with the mass $f = M_P{\widetilde f}$, with ${\widetilde f} \equiv 4\pi(1 - b)$. Therefore, the $b \rightarrow 0$ limit corresponds to the source with the critical mass $f_c = 4\pi M_P$ (for which the deficit angle near the origin, as in the massless and linearized massive cases, is $\theta = 2\pi$).

{}Recall that at large $\sigma$ we have $y\sim ({\widetilde f}_1 /2\pi)K_0(\sigma)$. In the linearized case ${\widetilde f}_1$ and ${\widetilde f}$ are the same. However, in the nonlinear case ${\widetilde f}_1$ is nontrivially related to both ${\widetilde f}$ (which is fixed by $b$) and $\zeta_*$.

\section{Codimension-2 Brane World}

{}In the previous section we discussed non-perturbative static solutions in $D=3$, where we have a point-like source at the origin, the space near the origin is a 2-dimensional ``wedge" with a deficit angle, and at distances much larger than the inverse perturbative graviton mass the space is flat with no deficit angle. Such solutions actually exist for all $D$.

{}Thus, consider a delta-function-like $p$-brane with $p=(D-3)$ and the brane tension $f\geq 0$ in $D$-dimensional bulk. The source term reads
\begin{equation}\label{source.D}
 S_{source} = - f \int_\Sigma d^{D-2}x \sqrt{-{\widehat G}}~,
\end{equation}
where the integral is over the worldvolume $\Sigma$, and ${\widehat G}_{\mu\nu} \equiv {\delta_\mu}^M {\delta_\nu}^N G_{MN}|_\Sigma$, where $x^\mu$, $\mu = 0,1,\dots,(D-3)$ are the coordinates on the brane, and the transverse coordinates are $x^i$, $i=1,2$: $x^M = (x^\mu, x^i)$. The bulk action is the same as (\ref{MG.action}). Then we have a static solution with the metric
\begin{equation}\label{metric.D}
 ds^2 = \eta_{\mu\nu} dx^\mu dx^\nu + \exp(2\omega)\delta_{ij} dx^i dx^j~,
\end{equation}
where $\omega$ is given by exactly the same non-perturbative solution to (\ref{eq.omega}) we discussed in Section 2 with the only difference that we now have
\begin{equation}
 {\widetilde f} = f / M_P^{D-2}~.
\end{equation}
So, we have a flat brane (with the Minkowski metric $\eta_{\mu\nu}$ on the brane) for a continuous range of the brane tension $0\leq f < f_c\equiv 4\pi M_P^{D-2}$. Gravity on the brane is not $D$-dimensional but $(D-2)$-dimensional -- once we add the Einstein-Hilbert term on the brane, that is (see Section 5 for details):
\begin{equation}\label{EH.brane}
 S_{brane} = {\widehat M}_P^{D-4}\int_\Sigma d^{D-2}x\sqrt{-{\widehat G}}{\widehat R}~,
\end{equation}
where ${\widehat M}_P$ is the $(D-2)$-dimensional Planck scale and ${\widehat R}$ is the scalar curvature on the brane constructed from the metric ${\widehat G}_{\mu\nu}$. The $(D-2)$-dimensional Einstein-Hilbert term can be induced on the brane via loop corrections so long as the matter on the brane is not conformal \cite{DGP, DG}. Alternatively, the Einstein-Hilbert term on a positive-tension brane can effectively be present classically if we include
higher curvature ({\em e.g.}, Gauss-Bonnet) terms in the bulk \cite{6D2}.

{}Finally, let us comment on higher-codimension $d \geq 3$ cases. From (\ref{R.eom}) we have the scalar curvature $R = \mu^2\left[G^{MN}\eta_{MN} - D\right]$. One can show (as in \cite{ZK2}) that, with massive gravity in the bulk, $d \geq 3$ solutions with positive brane tension are always singular -- unless one includes higher curvature terms as in \cite{CIK}. Then one expects to have flat brane solutions for a continuous range of the brane tension as in \cite{CIK}.

\section{String Solutions}

{}In Section 2 we discussed static solutions for point-like sources in $D=3$, which in higher dimensions generalize to codimension-2 branes. In this section we will first discuss string solutions in $D=3$ and then their generalizations to higher dimensions.

\subsection{Massless Case}

{}As in Section 2, let us start with the massless case, which we discussed in Subsection \ref{sub.massless}. There we found a solution with a point-like source term at the origin. There also exist solutions with sting-like sources. Thus, consider the following solution:
\begin{eqnarray}
 &&\omega = -{1\over 4\pi} {\widetilde f}~\ln\left(\rho\over \rho_0\right)~,~~~\rho > \rho_0~,\\
 &&\omega = 0~,~~~\rho\leq \rho_0~,
\end{eqnarray}
where $\rho_0$ is a parameter. This solution satisfies the following equation (the $(\rho,\phi)$ coordinates are defined in Subsection \ref{sub.massless}):
\begin{equation}
 \partial^i\partial_i \omega = -{1\over 4\pi\rho_0} {\widetilde f}~\delta(\rho - \rho_0)~.
\end{equation}
This implies that we have a closed circular string of radius $\rho_0$ with its center at the origin. The corresponding source term in the action reads:
\begin{equation}\label{string.source}
 S_{source} = - f \int_\Sigma d^2 x \sqrt{-\widehat G}~,
\end{equation}
where the worldvolume $\Sigma = R^1 \times S^1$ ($R^1$ corresponds to time $t$, and $S^1$ has radius $\rho_0$), and ${\widehat G}_{ab} \equiv {\delta_a}^M {\delta_b}^N G_{MN}|_\Sigma$, where $x^a \equiv (t, \rho_0\phi)$. The string tension $f \equiv M_P {\widetilde f}/2\pi\rho_0$. For $\rho > \rho_0$ the space is flat but has a deficit angle, and we have $0\leq f < f_c$, where $f_c\equiv 2M_P/\rho_0$. For $\rho\leq \rho_0$ the space is flat with no deficit angle.

\subsection{Massive Case}

{}In the massive case, in the presence of the string source term (\ref{string.source}), the full non-perturbative equation of motion for $\omega$ reads:
\begin{equation}\label{massive.eq}
 \partial^i\partial_i \omega + \mu^2\left[1 - \exp(2\omega)\right] = -{1\over 4\pi\rho_0} {\widetilde f}~\delta(\rho - \rho_0)~,
\end{equation}
where, as in the massless case, ${\widetilde f}\equiv 2\pi\rho_0 f / M_P$. As in the case of a point-like source, we will analyze this equation in two steps.

\subsection{Linearized Approximation}

{}As we will see below, for small brane tension the linearized approximation is actually valid everywhere. The linearized equation reads:
\begin{equation}
 \partial_\sigma^2 z + {1\over \sigma}\partial_\sigma z - z = -{1\over 2\pi\sigma_0} {\widetilde f}~\delta(\sigma - \sigma_0)~,
\end{equation}
where $z \equiv 2\omega$, $\sigma\equiv M\rho$, $\sigma_0\equiv M\rho_0$ and $M\equiv\sqrt{2}\mu$. The solution is given by
\begin{eqnarray}
 && z = {{\widetilde f} \over 2\pi}~{I_0(\sigma_0)\over Q}~K_0(\sigma)~,~~~\sigma > \sigma_0~,\\
 && z = {{\widetilde f} \over 2\pi}~{K_0(\sigma_0)\over Q}~I_0(\sigma)~,~~~\sigma \leq \sigma_0~,
\end{eqnarray}
where $Q\equiv \sigma_0\left[I_0^\prime(\sigma_0) K_0(\sigma_0) - K_0^\prime(\sigma_0) I_0(\sigma_0)\right]$. Note that $I_0(0)=1$ and for small $\sigma$ we have $I^\prime_0(\sigma)\sim \sigma/2$, $K_0(\sigma)\sim -\ln(\sigma)$, and $K_0^\prime(\sigma) \sim -1/\sigma$, so $Q \rightarrow 1$ for $\sigma_0\rightarrow 0$. If ${\widetilde f}\ll 1$, then the linearized approximation is valid as $z$ is finite and small for all $\sigma$. This is to be contrasted with the point-like source case, where the linearized approximation breaks down near the origin, where the solution blows up.

\subsection{Non-perturbative Solutions}

{}Solutions to the full non-perturbative equation (\ref{massive.eq}) have a structure similar to the linearized case. There are two independent solutions to equation (\ref{eq.y}) without the source term. One, call it ${\widetilde K}(\sigma)$, blows up at the origin and decays to zero at infinity. Another, call it ${\widetilde I}(\sigma)$, blows up at infinity and is finite at the origin. The solution to the full non-perturbative equation (\ref{massive.eq}) then reads:
\begin{eqnarray}
 && y = {{\widetilde f} \over 2\pi}~{{\widetilde I}(\sigma_0)\over {\widetilde Q}}~{\widetilde K}(\sigma)~,~~~\sigma > \sigma_0~,\\
 && y = {{\widetilde f} \over 2\pi}~{{\widetilde K}(\sigma_0)\over {\widetilde Q}}~{\widetilde I}(\sigma)~,~~~\sigma \leq \sigma_0~,
\end{eqnarray}
where ${\widetilde Q}\equiv \sigma_0\left[{\widetilde I}^\prime(\sigma_0) {\widetilde K}(\sigma_0) - {\widetilde K}^\prime(\sigma_0) {\widetilde I}(\sigma_0)\right]$, and $y\equiv 2\omega$.

{}We discussed how to construct ${\widetilde K}(\sigma)$ numerically in Subsection \ref{nonpert}. Similarly, we can also construct ${\widetilde I}(\sigma)$ numerically by setting ${\widetilde I}(0) = {\widetilde f}_2/2\pi$ and ${\widetilde I}^\prime(0) = 0$, where ${\widetilde f}_2$ is a parameter. The second-order ordinary differential equation (\ref{eq.y}) can then be solved for ${\widetilde I}(\sigma)$ using the standard numeric methods (Euler, Runge-Kutta, {\em etc.}).

\subsection{Codimension-2 Brane on a String}

{}Earlier in this section we discussed string solutions in $D=3$. Such solutions actually exist for all $D$. We have a brane with a worldvolume $\Sigma = R^{1,D-3} \times S^1$ embedded in $D$ dimensions. The source term reads
\begin{equation}
 S_{source} = - f \int_\Sigma d^{D-1}x \sqrt{-{\widehat G}}~,
\end{equation}
where ${\widehat G}_{ab} \equiv {\delta_a}^M {\delta_b}^N G_{MN}|_\Sigma$, $x^a \equiv (x^\mu, \rho_0\phi)$, and the $x^\mu$, $\mu = 0,1,\dots,(D-3)$, correspond to $R^{1,D-3}$. The metric is given by
\begin{equation}
 ds^2 = \eta_{\mu\nu} dx^\mu dx^\nu + \exp(2\omega)\delta_{ij} dx^i dx^j~,
\end{equation}
where $\omega$ is given by exactly the same non-perturbative solution to (\ref{massive.eq}) we discussed earlier in this section in the $D=3$ case with the only difference that we now have ${\widetilde f} = 2\pi\rho_0 f / M_P^{D-2}$. As in Section 3, we can include the Einstein-Hilbert term on the brane and obtain $(D-2)$-dimensional gravity in the noncompact part of the brane worldvolume (see Section 5 for details).\footnote{\, Depending on a detailed structure of the gravitational part of the action on the brane, gravity can be $(D-1)$-dimensional at distance scales below $\rho_0$, which is assumed to be suitably small.}

{}Finally, let us comment on higher-codimension $d\geq 3$ cases. As in \cite{CIK}, the brane worldvolume can be taken to be $R^{1,D-d-1}\times S^{d-1}$. Once higher curvature terms in the bulk are included, we expect to have nonsingular solutions as in \cite{CIK}.

\section{Brane Gravity}

{}In this section we discuss gravity in the Brane World (\ref{BW}) in the presence of matter fields on the brane. To do this, let us first discuss coupling of massive bulk gravity to a general matter configuration:
\begin{equation}\label{matter}
 S = M_P^{D-2} \int d^D x \sqrt{-G}\left[R - \mu^2 V\right] + S_{matter}~.
\end{equation}
We will assume that the energy-momentum tensor defined as
\begin{equation}
 T_{MN}\equiv 2 ~{\delta S_{matter} \over \delta G^{MN}}
\end{equation}
is conserved:
\begin{equation}\label{EM}
 \nabla^M T_{MN} = 0.
\end{equation}
This is because in the gravitational Higgs mechanism (which we are assuming here as the origin of the ``mass" term $\mu^2 V$ in (\ref{matter})) diffeomorphisms are broken spontaneously, and the $D$ scalar fields that give rise to $V$ are all set to their background values, so they do not contribute to $T_{MN}$. The matter can (but does not have to be) localized on a brane. In this case, for the sake of notational convenience we will include the contributions from the Einstein-Hilbert term on the brane (as in (\ref{BW})) into the definition of $S_{matter}$ (see below).

{}For our purposes here it will suffice to consider the cases where $V$ is a function of $X \equiv G^{MN}{\widetilde G}_{MN}$, where ${\widetilde G}_{MN}$ is a solution to the equations of motion without the matter fields. Because of (\ref{EM}), even in the presence of the matter fields, we have the Bianchi identity:
\begin{equation}\label{Bianchi.1}
 \partial_M\left[\sqrt{-G}V^\prime(X)G^{MN}{\widetilde G}_{NS}\right] - {1\over 2}\sqrt{-G}V^\prime(X)G^{MN}\partial_S{\widetilde G}_{MN} = 0~.
\end{equation}
In the following we will focus on the cases with the flat Minkowski background metric ${\widetilde G}_{MN} = \eta_{MN}$. This requires that
\begin{equation}
 V(D) = 2V^\prime(D)~.
\end{equation}
We then have
\begin{equation}\label{Bianchi.2}
 \partial_M\left[\sqrt{-G}V^\prime(X)G^{MN}\right] = 0~,
\end{equation}
which follows from (\ref{Bianchi.1}).

\subsection{Linearized Analysis}

{}Let $G_{MN}\equiv \eta_{MN} + h_{MN}$ and $h\equiv \eta^{MN}h_{MN}$. Then, in the linearized approximation, we have the following perturbative graviton mass term in the action
\begin{equation}
 -{M^2\over 4}\left[h_{MN}h^{MN} - \beta~h^2\right]~,
\end{equation}
where
\begin{eqnarray}
 &&M^2\equiv 2\mu^2 V^\prime(D)~,\\
 &&\beta \equiv {1\over 2} - {V^{\prime\prime}(D) \over {V^\prime(D)}}~.
\end{eqnarray}
When $V^{\prime}(D) = -2 V^{\prime\prime}(D)$, we have $\beta=1$ and the Fierz-Pauli mass term. However, we will keep $\beta$ arbitrary for now. The linearized Bianchi identity (\ref{Bianchi.2}) reads:
\begin{equation}
 \partial^M h_{MN} = \beta \partial_N h~,
\end{equation}
and the linearized equations of motion are given by:
\begin{eqnarray}
 &&-\partial^L\partial_L h_{MN} + (2\beta-1)\partial_M\partial_N h + (1 - \beta)\eta_{MN}\partial^L\partial_L h + \nonumber\\
 &&\,\,\,\,\,\,\, + M^2\left[h_{MN} - \beta \eta_{MN} h\right] =
 M_P^{2-D}~T_{MN}~.\label{EOMhMN}
\end{eqnarray}
The energy-momentum tensor conservation condition reads $\partial^M T_{MN} = 0$. The trace equation of motion reads:
\begin{equation}\label{EOMh}
 -(D-2)(1-\beta) \partial^L\partial_L h + (D\beta - 1) M^2 h = -M_P^{2 - D}~T~,
\end{equation}
where $T \equiv \eta^{MN} T_{MN}$. The perturbative mass squared for the traceless part of $h_{MN}$ is $M^2$, while for the trace $h$ it is given by $M_h^2 \equiv (D\beta - 1) M^2 / (D-2)(1-\beta)$. When $\beta\rightarrow 1-$, we have $M_h^2\rightarrow \infty$, so {\em naively} the trace $h$ decouples. However, at $\beta = 1$ we have nonzero $h$ so long as $T$ is nonzero, and if $T$ is singular (see below), so is $h$, which means that the perturbative approximation breaks down altogether. Another ``special" point is $\beta = 1/2$, where the $\partial_M\partial_N h$ term in (\ref{EOMhMN}) vanishes. This term becomes important when we consider singular matter sources, that is, lower-dimensional sources of nonzero codimension $d$.

{}Thus, consider matter localized on a hypersurface $\Sigma$ of codimension $d\geq 1$, {\em i.e.}, $\Sigma$ is a $p$-brane with $p = D - d - 1$. Let the coordinates transverse to $\Sigma$ be $x^i$, while the coordinates along $\Sigma$ be $x^\mu$. Then we have $T_{\mu i} = 0$, $T_{ii} = 0$, $T = \eta^{\mu\nu} T_{\mu\nu}$, and
\begin{eqnarray}
 &&-\partial^i \partial_i h_{\mu\nu} - \partial^\lambda\partial_\lambda h_{\mu\nu} + M^2h_{\mu\nu} + (2\beta-1)\left[\partial_\mu\partial_\nu h + \eta_{\mu\nu} M^2 h\right] = \nonumber\\
 &&\,\,\,\,\,\,\,M_P^{2-D}\left[T_{\mu\nu} - {1\over{D-2}}\eta_{\mu\nu} T\right]~.
\end{eqnarray}
Let us Fourier transform the coordinates $x^\mu$ on the brane and rotate to the Euclidean space. We have
\begin{eqnarray}
 &&(D-2)(1-\beta) \left[-\partial^i\partial_i h + p^2 h\right] + (D\beta - 1) M^2 h = \nonumber\\
 &&\,\,\,\,\,\,\,-M_P^{2 - D}~{\widehat T}\delta^{(d)}(x^i - x^i_*)~,\label{F-h}\\
 &&-\partial^i \partial_i h_{\mu\nu} + \left(p^2 + M^2\right) h_{\mu\nu} + (2\beta-1)\left[\eta_{\mu\nu} M^2 - p_\mu p_\nu \right] h = \nonumber\\
 &&\,\,\,\,\,\,\,M_P^{2-D}\left[{\widehat T}_{\mu\nu} - {1\over{D-2}}\eta_{\mu\nu} {\widehat T}\right] \delta^{(d)}(x^i - x^i_*)~,\label{F-hmunu}
\end{eqnarray}
where $x^i_*$ are the coordinates of the location of the brane $\Sigma$ in the transverse space, ${\widehat T}_{\mu\nu}$ is independent of $x^i$ ({\em i.e.}, it is the energy-momentum tensor on the brane), ${\widehat T}\equiv \eta^{\mu\nu}{\widehat T}_{\mu\nu}$, $p^\mu {\widehat T}_{\mu\nu} = 0$, and ${\widehat T}_{\mu\nu}$ includes the contribution from the Einstein-Hilbert term on the brane:
\begin{equation}
 S_{matter} \equiv {\cal S}_{matter} + {\widehat M}_P^{D-d-2}\int_\Sigma dx^{D-d} \sqrt{-{\widehat G}} {\widehat R}~,
\end{equation}
where ${\cal S}_{matter}$ is the pure brane matter contribution. We therefore have
\begin{equation}\label{calT}
 {\widehat T}_{\mu\nu} = {\cal T}_{\mu\nu} + {\widehat M}_P^{D-d-2}\sqrt{-{\widehat G}}\left[ {\widehat G}_{\mu\nu} {\widehat R} - 2 {\widehat R}_{\mu\nu} \right]~,
\end{equation}
where ${\cal T}_{\mu\nu} \equiv 2\delta{\cal S}_{matter}/\delta {\widehat G}_{\mu\nu}$, ${\widehat G}_{\mu\nu} \equiv {\delta_\mu}^M{\delta_\nu}^N G_{MN}|_\Sigma$, and the $(D-d)$-dimensional scalar curvature ${\widehat R}$ and the Ricci tensor ${\widehat R}_{\mu\nu}$ on the brane are constructed from ${\widehat G}_{\mu\nu}$. In the linearized approximation, after the Fourier transformation, we have
\begin{equation}
 {\widehat T}_{\mu\nu} = {\cal T}_{\mu\nu} - {\widehat M}_P^{D-d-2}\left[p^2H_{\mu\nu} - 2p^\lambda p_{(\mu} H_{\nu)\lambda} +
 p_\mu p_\nu H - \eta_{\mu\nu}\left(p^2 H - p^\lambda p^\sigma H_{\lambda\sigma}\right)\right]~.
\end{equation}
The $d=1$ and $d>1$ cases require separate treatment as in $d>1$ we must smooth out the singularity at the origin.

\subsubsection{$d = 1$}

{}Let the transverse coordinate be $y$. Then we have
\begin{equation}
 h = -{M_P^{2-D}{\widehat T}\over(D-2)(1-\beta)}~{\exp\left(-\sqrt{p^2 + M_h^2}|y|\right)\over 2\sqrt{p^2 + M_h^2}}~,
\end{equation}
and
\begin{eqnarray}
 &&h_{\mu\nu} = M_P^{2-D}\left[{\widehat T}_{\mu\nu} - {1\over{D-2}}\eta_{\mu\nu} {\widehat T}\right]{\exp\left(-\sqrt{p^2 + M^2}|y|\right)\over 2\sqrt{p^2 + M^2}} - {M_P^{2-D} {\widehat T}\over {(D-1)M^2}}\times \nonumber\\
 &&\,\,\,\, \times \left(\eta_{\mu\nu}M^2 - p_\mu p_\nu\right)\left[{\exp\left(-\sqrt{p^2 + M_h^2}|y|\right)\over 2\sqrt{p^2 + M_h^2}} - {\exp\left(-\sqrt{p^2 + M^2}|y|\right)\over 2\sqrt{p^2 + M^2}}\right],
\end{eqnarray}
where we have used $M_h^2 - M^2 = (D-1)(2\beta-1)M^2/(D-2)(1-\beta)$. Let $H_{\mu\nu}\equiv h_{\mu\nu}(y = 0)$, and $H\equiv \eta^{\mu\nu}H_{\mu\nu}$. On the brane we have:
\begin{eqnarray}
 &&H_{\mu\nu} = M_P^{2-D}\left[{\widehat T}_{\mu\nu} - {1\over{D-2}}\eta_{\mu\nu} {\widehat T}\right]{1\over 2\sqrt{p^2 + M^2}} - {M_P^{2-D} {\widehat T}\over {(D-1)M^2}}\times \nonumber\\
 &&\,\,\,\, \times \left(\eta_{\mu\nu}M^2 - p_\mu p_\nu\right) \left[{1\over 2\sqrt{p^2 + M_h^2}} - {1\over 2\sqrt{p^2 + M^2}}\right],\label{Hmunu}
\end{eqnarray}
and
\begin{equation}\label{H}
 H = -{M_P^{2-D}{\widehat T}\over 2\sqrt{p^2 + M^2}} \left({1\over{D-2}} +
 {{(D-1)M^2 - p^2}\over{(D-1)M^2}}\left[\sqrt{{p^2 + M^2}\over {p^2 + M_h^2}} - 1\right]\right).
\end{equation}
Recall that ${\widehat T}_{\mu\nu}$ itself depends on $H_{\mu\nu}$ via (\ref{calT}). Our goal is to solve for $H_{\mu\nu}$ via ${\cal T}_{\mu\nu}$, which is the energy-momentum tensor of the pure brane matter. To do this, let us note that
\begin{equation}
 {\widehat T} = {\cal T} - {\widehat M}_P^{D-3}(D-3)\left[p^\mu p^\nu H_{\mu\nu} - p^2 H\right],
\end{equation}
where ${\cal T}\equiv \eta^{\mu\nu} {\cal T}_{\mu\nu}$, and
\begin{equation}\label{pH}
 p^\mu H_{\mu\nu} - p_\nu H = p_\nu~{M_P^{2-D}{\widehat T}\over 2\sqrt{p^2 + M^2}} {{D-2}\over{D-1}} \left[\sqrt{{p^2 + M^2}\over {p^2 + M_h^2}} - 1\right].
\end{equation}
The $\beta = 1/2$ and $\beta\not=1/2$ cases require separate treatment.

\medskip
{\em $\bullet$ The $\beta = 1/2$ Case}
\medskip

{}In this case we have $M_h = M$, $p^\mu H_{\mu\nu} = p_\nu H$, ${\widehat T} = {\cal T}$, and
\begin{eqnarray}
 &&H_{\mu\nu} = {{\widehat M}_P^{3-D}\over {p^2 + m_*\sqrt{p^2 + M^2}}}~\left[{\cal T}_{\mu\nu} -
 {{\cal T}\over{D-2}}\left(\eta_{\mu\nu} + {p_\mu p_\nu \over m_*\sqrt{p^2 + M^2}}\right)\right],\\
 &&H = -{M_P^{2-D}{\cal T}\over 2\sqrt{p^2 + M^2}(D-2)}~,\label{H-beta1/2}
\end{eqnarray}
where $m_* \equiv 2M_P^{D-2} / {\widehat M}_P^{D-3}$. Note that in the massless case $r_* \equiv 1/m_*$ plays the role of the cross-over scale below which gravity on the brane is $(D-1)$-dimensional, while above this scale it becomes $D$-dimensional. However, in the massive case the situation is different. Assuming the graviton mass $M\gg m_*$, at momenta $p^2 \ll M^2$ the graviton propagator on the brane behaves as $1/(p^2 + m_c^2)$, where $m^2_c \equiv m_* M$. The actual ``cross-over" scale, therefore, is:
\begin{equation}
 r_c \equiv 1/m_c = \sqrt{{\widehat M}_P^{D-3} \over 2M_P^{D-2}M}~.
\end{equation}
In $D=5$ ({\em i.e.}, in the case of a 3-brane in 5 dimensions) we can have the bulk graviton mass as high as the bulk Planck scale, $M\sim M_P$, in which case the ``cross-over" scale
\begin{equation}
 r_c \sim {{\widehat M}_P\over 2M_P^2}~,
\end{equation}
which is of order of the current Hubble size when $M_P \sim (0.1~{\mbox{mm}})^{-1}$. Note, however, that just as in the massless case \cite{DGP}, here too the tensor structure of the graviton propagator is $D$-dimensional and not $(D-1)$-dimensional in that it couples to ${\cal T}_{\mu\nu} - \eta_{\mu\nu}{\cal T} / (D-2)$ and not to ${\cal T}_{\mu\nu} - \eta_{\mu\nu}{\cal T} / (D-3)$. As we will see in a moment, this is not the case for other values of $\beta$. We also note that the term proportional to $p_\mu p_\nu$ in the graviton propagator does not affect the gravitational coupling on the brane as is does not couple to the conserved energy-momentum tensor on the brane, albeit {\em a priori} it does couple to the bulk matter. It is this term that causes $H$ in (\ref{H-beta1/2}) to be purely $D$-dimensional, which is suppressed exponentially at distances $r\gg 1/M$ because bulk gravity is massive. Finally, we have been using quotation marks when referring to the ``cross-over" scale $r_c$ because, unlike in the massless case, at distances $r\gsim r_c$ gravity actually does not become $D$-dimensional. Instead, the momentum structure of the graviton propagator on the brane has the form of a massive propagator with mass squared $m_c^2$.

\medskip
{\em $\bullet$ The $\beta \not= 1/2$ Case}
\medskip

{}In this case, we can solve for the graviton propagator on the brane as follows. We have the following general tensor structure:
\begin{equation}\label{tensor}
 H_{\mu\nu} = a{\cal T}_{\mu\nu} + b\eta_{\mu\nu}{\cal T} + c p_\mu p_\nu {\cal T}~.
\end{equation}
The value of $a$ is evident from (\ref{Hmunu}) and (\ref{calT}):
\begin{equation}
 a = {{\widehat M}_P^{3-D}\over {p^2 + m_*\sqrt{p^2 + M^2}}}~.
\end{equation}
It is independent of $\beta$. The value of $b$ can be deduced by noting that
\begin{eqnarray}
 &&p^\mu H_{\mu\nu} - p_\nu H = -p_\nu\left[a + (D-2)b\right]{\cal T}~,\\
 &&{\widehat T} = {\cal T} \left(1 + {\widehat M}_P^{D-3}(D-3)p^2\left[a + (D-2)b\right]\right).\label{hatT}
\end{eqnarray}
Using (\ref{pH}) we have
\begin{eqnarray}
 &&b = -{{\widehat M}_P^{3-D}\over {D-2}} \left(p^2 + m_*\sqrt{p^2 + M^2}\right)^{-1} - \nonumber\\
 &&\,\,\,{{\widehat M}_P^{3-D}\over {(D-2)(D-3)}}\left(p^2 +
 {m_*(D-1)\over (D-2)(D-3)}{\sqrt{p^2 + M^2}\sqrt{p^2 + M_h^2} \over {\sqrt{p^2 + M^2} - \sqrt{p^2 + M_h^2}}}\right)^{-1}.\label{b}
\end{eqnarray}
Recall that $M_h^2/M^2 = (D\beta - 1)/(D-2)(1-\beta)$, so $M_h > M$ for $\beta > 1/2$, $M_h < M$ for $\beta < 1/2$, and $M_h = M$ for $\beta = 1/2$. Therefore, the allowed values of $\beta$ are between $1/D$ and 1, and for $\beta>1/2$ the second set of round brackets in (\ref{b}) may have a pole, which would imply presence of a tachyon. Let us therefore discuss the $\beta>1/2$ and $\beta<1/2$ cases separately.

\medskip
{\em $\bullet$ The $\beta < 1/2$ Case}
\medskip

{}In this case there is no tachyon pole. Assuming $m_* \ll M \sim M_P$, $m_* \ll M_h$ and $1 - M_h / M$ is not small, we have for $p^2 \ll M^2$:
\begin{equation}
 b\approx -{{\widehat M}_P^{3-D}\over(D-2)(D-3)}\left[{{D-3}\over{p^2 + m_c^2}} + {1\over {p^2 + m_1^2}}\right],
\end{equation}
where $m_1^2\equiv m_c^2 (D-1)M_h/(D-2)(D-3)(M-M_h)\sim m_c^2$. This implies that at $p^2\gg m_c^2$ we have $b\approx -{\widehat M}_P^{3-D}/(D-3)$, {\em i.e.}, we have the $(D-1)$-dimensional tensor structure:
\begin{equation}
 H_{\mu\nu} \approx {{\widehat M}_P^{3-D}\over p^2}\left[{\cal T}_{\mu\nu} - {1\over{D-3}}\eta_{\mu\nu}{\cal T}\right]+\dots~,
\end{equation}
where the ellipses include the terms proportional to $p_\mu p_\nu$ along with the subleading terms. We do, however, need to make sure that the terms proportional to $p_\mu p_\nu$ do not have a tachyon pole, {\em i.e.}, we need to show that $c$ in (\ref{tensor}) has no tachyon pole.

{}To see that this is indeed the case, note from (\ref{tensor}) that $H = (a + (D-1)b + cp^2){\cal T}$. This implies that, since $a$ has no pole and $b$ has no pole, then $c$ has no pole so long as $H$ has no pole. From (\ref{H}) it follows that $H$ has no pole so long as ${\widehat T}$ has no pole. On the other hand, (\ref{hatT}) implies that ${\widehat T}$ has no pole since neither $a$ nor $b$ does.

{}If we assume that $M - M_h \ll M$, then $m_1^2 \gg m_c^2$, so the tensor structure becomes $D$-dimensional at momenta $p^2\lsim m_1^2$. In fact, if we take the limit $\beta\rightarrow 1/2-$ ($M_h\rightarrow M-$), then we have $D$-dimensional tensor structure at all momenta, which is consistent with what we found above for $\beta = 1/2$. However, we can relax the assumption that $m_* \ll M_h$. In fact, $M_h$ can be as small as 0. {\em E.g.}, if $M_h = 0$, we have (at $p^2 \ll M^2$):
\begin{equation}
 b\approx -{{\widehat M}_P^{3-D}\over(D-2)(D-3)}\left[{{D-3}\over{p^2 + m_c^2}} + {1\over{p^2 + m_2 p}}\right],
\end{equation}
where $p\equiv\sqrt{p^2}$ and $m_2\equiv m_*(D-1)/(D-2)(D-3)$. Since $m_2 \sim m_c^2 / M_P \ll m_c$, for $p^2\gg m_c^2$ we have the $(D-1)$-dimensional tensor structure.

\medskip
{\em $\bullet$ The $\beta > 1/2$ Case}
\medskip

{}Assuming $\beta > 1/2$, $m_* \ll M$ and $m_*\ll M_h$, we indeed have a tachyon pole at $p^2 = m_T^2$, where $m_T^2 \approx m_* M M_h (D-1)/(D-2)(D-3)(M_h - M)\sim m_c^2$ -- assuming $M_h - M$ is not small, that is (see below). Even if we take $\beta \rightarrow 1-$, we still have a tachyon pole with $m_T^2\approx m_c^2 (D-1)/(D-2)(D-3)$. In fact, as mentioned above, in this limit the trace $h$ goes to infinity on the brane, so the perturbative expansion breaks down. Note that $H$ is finite in this limit, so it is actually the radion ({\em i.e.}, the transverse component $h_{DD}$) that diverges on the brane.

{}Here we can ask: Is the tachyon with $m_T^2\sim m_c^2$ really bad? The presence of a tachyon with $m_T^2\sim m_c^2$ does not necessarily spell a disaster. First, $m_T\gsim H$, so the instability due to the tachyon is not observable. Second, as was pointed out in \cite{Orientiworld}, the presence of a ``light" tachyon of this kind is likely an artifact of dropping an infinite tower of nonlocal terms on the brane suppressed by powers of $m_c^2/p^2$. Our treatment is only valid in the range $m_c^2\lsim p^2\lsim M_P^2$. In fact, a tachyon pole $(p^2 - m_T^2)^{-1}$, when expanded in powers of $m_c^2/p^2$ (assuming that $m_T\sim m_c$), would give us a clue about some nonlocal terms that must be included on the brane. These nonlocal terms can be safely ignored when $p^2\gg m_c^2$, but become important when $p^2\lsim m_c^2$, which is precisely when the tachyon pole becomes relevant.

{}Let us now turn to the case where $M_h > M$ but $M_h - M\lsim m_*$, then there need not be a tachyon pole. In fact, we expect this to be the case on the grounds that for $\beta=1/2$ we do not have a tachyon. Let us look into this in more detail. Let $M_h = M(1 + \epsilon)$, where $0 < \epsilon \ll 1$. Then a tachyon pole in (\ref{b}) occurs at $p^2 = m_T^2 \equiv M^2 \zeta$, where $\zeta$ is a root of the equation $\zeta - \kappa (\zeta + 1)^{3/2} = 0$, where $\kappa\approx(m_* / \epsilon M)(D-1)/(D-2)(D-3)$. Actually, this equation has two roots when $\kappa < \kappa_*$, one root when $\kappa = \kappa_*$, and no roots when $\kappa > \kappa_*$, where $\kappa_* = (2/3)^{3/2}$. If $\kappa \ll 1$, then $\zeta\approx\kappa$, and $m_T^2 \approx m_c^2 (D-1)/(D-2)(D-3)\epsilon \gg m_c^2$, so the pole becomes more bothersome.\footnote{\,The other root is $\zeta\approx 1/\kappa^2\gg 1$, which is physically irrelevant as the low energy field theory approximation breaks down at momenta $p^2\gsim M_P^2$ and we are assuming $M\sim M_P$.} In particular, our approximation (which includes linearization and neglecting nonlocal terms) can no longer be trusted for momenta $p^2\lsim m_c^2 / \epsilon$. If $\kappa \leq \kappa_*$ and $\kappa\sim 1$, then the tachyon pole(s) with $m_T \sim M_P$ is (are) irrelevant (as the effective field theory is valid at $p^2\ll M_P^2$). If $\kappa > \kappa_*$, which occurs when $\epsilon < \epsilon_*$, where $\epsilon_* \approx (m_* / M)(D-1)/(D-2)(D-3)\kappa_*$, then, consistently with the $\beta=1/2$ case, there is no tachyon pole. Either way, the tensor structure is $D$-dimensional.

{}If we can ignore the tachyon pole for the reasons mentioned above -- assuming $m_T\sim m_c$, that is -- then we have the $(D-1)$-dimensional tensor structure in the graviton propagator for all the corresponding values of $\beta$ as at $p^2\gg m_c^2$ we have $b\approx -{\widehat M}_P^{3-D}/(D-3)p^2$ -- see (\ref{b}).

\subsubsection{Why Is the Tensor Structure $(D-1)$-dimensional?}

{}Here it appears that, in the codimension-1 case, massive gravity in the bulk achieves what massless gravity could not: In the latter case the tensor structure in the graviton propagator on the brane is always $D$-dimensional \cite{DGP}, unless one introduces a scalar ghost \cite{DGP1, GRS, CEH}. What plays the role of that scalar ``ghost" in the massive case is the trace $h$, which {\em perturbatively} is a ghost. However, the ``ghostliness" of $h$ is an artifact of linearization \cite{Unitarity}. The non-perturbative Hamiltonian for the relevant degrees of freedom (that is, the conformal and helicity-0 modes) is bounded from below even if $\beta\not=1$ ({\em i.e.}, when perturbatively the graviton mass term is not of the Fierz-Pauli form and perturbatively the theory is naively non-unitary) \cite{Unitarity}. The full non-perturbative Hamiltonian in the $\beta=1/2$ case was shown to be positive-definite in \cite{JK}. In fact, as we saw above, even in the $\beta \rightarrow 1$ case we have $(D-1)$-dimensional tensor structure. Naively, this might appear to make no sense as at $\beta=1$ perturbatively there is no ghost. However, as we saw above, the trace $h$ does not decouple from the brane matter sources even at $\beta=1$ and, in fact, the perturbative expansion breaks down altogether. Also, in the range of the values of $\beta$ near $1/2$ we found that there is no tachyon pole and the tensor structure is $D$-dimensional. This is consistent with the {\em perturbative} arguments of \cite{DGP, DGP1}, that some ``unconventional" states are needed for the tensor structure to be $(D-1)$-dimensional. However, the difference here is that massive gravity provides such ``unconventional" states in its naive linearization without violating unitarity as the theory is unitary non-perturbatively. Therefore, at least in the cases with $\beta < 1/2$ (no tachyon), with $\beta$ not too close to $1/2$ (see above), and possibly for other values of $\beta>1/2$ with the tachyon pole at $p^2 = m_T^2\lsim m_c^2$, we appear to have a consistent mechanism -- via massive gravity in the bulk -- for obtaining $(D-1)$-dimensional tensor structure on a codimension-1 brane in infinite-volume extra space.

\subsubsection{$d>1$}

{}When codimension $d>1$, we must regularize the $\delta$-function $\delta^{(d)}(x^i-x^i_*)$ in the source term. This regularization was discussed in detail in \cite{Orientiworld}. One replaces the codimension-$d$ $p$-brane with the worldvolume $R^{1,p}$ (where $p=D-d-1$) with a partially smoothed out brane whose worldvolume $\Sigma = R^{1,p} \times S^{d-1}$, where $S^{d-1}$ has a small radius $r_0$. The brane is now effectively of codimension 1, and the $d$-dimensional $\delta$-function $\delta^{(d)}(x^i-x^i_*)$ in the source term is replaced by a singular distribution $f(r) \equiv \delta(r- r_0)/a_{d-1}r_0^{d-1}$, where $a_{d-1}$ is the area of a $(d-1)$-sphere with unit radius, and $r$ is the radial coordinate in the transverse space. With this smoothing out, we can solve the equations of motion for (\ref{F-h}) and (\ref{F-hmunu}). The solutions are radially symmetric in the transverse space.

\medskip
{\em $\bullet$ The $d=3$ Case}
\medskip

{}For $d=3$ the solutions are elementary functions, so for the sake of simplicity we will focus on this case first for explicit computations and then give the answer for general $d\geq 3$. (The $d=2$ case requires separate treatment -- see below.) Thus, assuming $d=3$, let ${\widetilde h}_{\mu\nu}\equiv rh_{\mu\nu}$ and ${\widetilde h}\equiv rh$. Then we have the following equations:
\begin{eqnarray}
 &&(D-2)(1-\beta) \left[-\partial_r^2 {\widetilde h} + p^2 {\widetilde h}\right] + (D\beta - 1) M^2 {\widetilde h} = \nonumber\\
 &&\,\,\,\,\,\,\,-{M_P^{2 - D}\over a_{d-1}r_0^{d-2}}~{\widehat T}\delta(r- r_0)~,\label{F-h.1}\\
 &&-\partial_r^2 {\widetilde h}_{\mu\nu} + \left(p^2 + M^2\right) {\widetilde h}_{\mu\nu} + (2\beta-1)\left[\eta_{\mu\nu} M^2 - p_\mu p_\nu \right] {\widetilde h} = \nonumber\\
 &&\,\,\,\,\,\,\,{M_P^{2 - D}\over a_{d-1}r_0^{d-2}}\left[{\widehat T}_{\mu\nu} - {1\over{D-2}}\eta_{\mu\nu} {\widehat T}\right] \delta(r- r_0)~.\label{F-hmunu.1}
\end{eqnarray}
Let as define $\Phi(r, u)$ as follows:
\begin{eqnarray}
 &&\Phi(r \leq r_0, u) \equiv {M_P^{2 - D}\over a_{d-1}r_0^{d-2}}~{\exp\left(-\sqrt{p^2 + u^2}r_0\right)\over\sqrt{p^2+u^2}}~{\sinh\left(\sqrt{p^2 + u^2}r\right)\over r}~,\\
 &&\Phi(r > r_0, u) \equiv {M_P^{2 - D}\over a_{d-1}r_0^{d-2}}~{\sinh\left(\sqrt{p^2 + u^2}r_0\right)\over\sqrt{p^2+u^2}}~{\exp\left(-\sqrt{p^2 + u^2}r\right)\over r}~.
\end{eqnarray}
Then we have:
\begin{eqnarray}\label{h.d}
 &&h = -{{\widehat T}~\Phi(r, M_h)\over(D-2)(1-\beta)}~,\\
 &&h_{\mu\nu} = \left[{\widehat T}_{\mu\nu} - {1\over{D-2}}\eta_{\mu\nu} {\widehat T}\right]\Phi(r, M)-\nonumber\\
 &&\,\,\,\,\,\,\, {\left(\eta_{\mu\nu}M^2-p_\mu p_\nu\right){\widehat T}\over (D-1)M^2}\left[\Phi(r, M_h) - \Phi(r, M)\right]~.\label{hmunu.d}
\end{eqnarray}
As in the codimension-1 case, let us define $H_{\mu\nu}\equiv h_{\mu\nu}(r = r_0)$ and $H\equiv \eta^{\mu\nu} H_{\mu\nu}$. Then on the brane we have
\begin{eqnarray}
 &&H_{\mu\nu} = {M_P^{2-D}\over a_{d-1}r_0^{d-1}}\left[{\widehat T}_{\mu\nu} - {1\over{D-2}}\eta_{\mu\nu} {\widehat T}\right]{{1-e^{-2\sqrt{p^2+M^2}r_0}}\over 2\sqrt{p^2 + M^2}} - {M_P^{2-D}{\widehat T}\over a_{d-1}r_0^{d-1}} \times \nonumber\\
 &&\,\,\,\, \times {{\eta_{\mu\nu}M^2 - p_\mu p_\nu}\over {(D-1)M^2}} \left[{{1-e^{-2\sqrt{p^2+M_h^2}r_0}}\over 2\sqrt{p^2 + M_h^2}} - {{1-e^{-2\sqrt{p^2+M^2}r_0}}\over 2\sqrt{p^2 + M^2}}\right].\label{Hmunu-d}
\end{eqnarray}
We will assume that the brane ``thickness" $r_0$ is small: $M \ll 1/r_0$ and $M_h\ll 1/r_0$. Then we have (up to subleading terms suppressed by powers of $r_0$):
\begin{equation}
 H_{\mu\nu} \approx {M_P^{2-D}\over a_{d-1}r_0^{d-2}}\left[{\widehat T}_{\mu\nu} - {1\over{D-2}}\eta_{\mu\nu} {\widehat T}\right].
\end{equation}
Note that the term proportional to $(\eta_{\mu\nu}M^2 - p_\mu p_\nu)$ is subleading (as the leading contributions from the two exponents cancel out) and has been omitted. A little algebra gives us the following solution (up to subleading terms suppressed by powers of $r_0$):
\begin{eqnarray}
 && H_{\mu\nu} \approx {\widehat M}_P^{2+d-D}\left[{\cal T}_{\mu\nu} - {\eta_{\mu\nu}{\cal T}\over{D-d-1}}\right]{1\over{p^2 + m_c^2}} - \nonumber\\
 &&\,\,\,\,\,\,\,{{\widehat M}_P^{2+d-D}{\cal T}\over(D-d-1)(D-d-2)}~{1\over{p^2 - m_T^2}}~,
\end{eqnarray}
where
\begin{eqnarray}\label{mc3}
 &&m_c^2\equiv a_{d-1}r_0^{d-2}M_P^{D-2}/{\widehat M}_P^{D-d-2}~,\\
 &&m_T^2\equiv m_c^2(D-2)/(d-1)(D-d-2)~.\label{mT3}
\end{eqnarray}
This result is similar to that of \cite{Orientiworld} in the massless case with the difference that in the massive case the terms proportional to $p_\mu p_\nu {\cal T}$ are subleading. The above result holds for all allowed values of $\beta$. This is because the $\beta$-dependent contributions are subleading. Just as in the massless case, we have a tachyon pole at $p^2 = m_T^2\sim m_c^2$. As mentioned above, such a tachyon pole is not necessarily problematic as the corresponding instability is not observable and it is likely an artifact of neglecting nonlocal terms on the brane suppressed by powers of $m_c^2/p^2$, which become important in the infrared where the tachyon pole is relevant.

\medskip
{\em $\bullet$ The $d > 3$ Case}
\medskip

{}The formula (\ref{mc3}) applies to $d=3$. For general $d\geq 3$ we have
\begin{eqnarray}\label{mcd}
 &&m_c^2 = (d-2) a_{d-1}r_0^{d-2}M_P^{D-2}/{\widehat M}_P^{D-d-2}~,\\
 &&m_T^2\equiv m_c^2(D-2)/(d-1)(D-d-2)~.\label{mTd}
\end{eqnarray}
The extra factor of $(d-2)$ compared with (\ref{mc3}) comes from the fact that the solution at $r\rightarrow r_0+$ in $d\geq 3$ behaves as $1/r^{d-2}$. Note that for $d\geq 3$ the graviton mass does not enter (\ref{mcd}) or (\ref{mTd}). This is because for small $r_0$ the jump conditions on the first derivatives (due to the $\delta$-function in the source terms) at $r=r_0$ are determined by short distance dynamics, which is independent of $M$. In the $r_0\rightarrow 0$ limit we have $m_c\rightarrow 0$ and $m_T\rightarrow 0$. If we ignore the tachyon, at $p^2\gg m_c^2$ we have ($D-d)$-dimensional tensor structure as $H_{\mu\nu}$ couples to ${\cal T}_{\mu\nu} - \eta_{\mu\nu}{\cal T} / (D-d-2)$, so gravity is $(D-d)$-dimensional.

\medskip
{\em $\bullet$ The $d=2$ Case}
\medskip

{}In $d=2$ the situation is a bit trickier than in $d\geq 3$ because the first subleading terms in $r_0$ are only logarithmically suppressed. The solution is still given by (\ref{h.d}) and (\ref{hmunu.d}), but with $\Phi(r, u)$ now defined as
\begin{eqnarray}\label{Phi2-}
 &&\Phi(r \leq r_0, u) \equiv {M_P^{2 - D}\over a_{d-1}r_0^{d-2}}~{K_0(\sqrt{p^2 + u^2}r_0)\over Q(\sqrt{p^2 + u^2}r_0)}~I_0(\sqrt{p^2 + u^2}r)~,\\
 &&\Phi(r > r_0, u) \equiv {M_P^{2 - D}\over a_{d-1}r_0^{d-2}}~{I_0(\sqrt{p^2 + u^2}r_0)\over Q(\sqrt{p^2 + u^2}r_0)}~K_0(\sqrt{p^2 + u^2}r),\label{Phi2+}
\end{eqnarray}
where $Q(z)\equiv z\left[K_0(z)I_0^\prime(z)- I_0(z)K_0^\prime(z)\right]$. Assuming $r_0 \ll 1/M$ and $r_0\ll 1/M_h$, we have:
\begin{eqnarray}
 &&H_{\mu\nu} \approx {M_P^{2-D}\over a_{d-1}r_0^{d-2}}\left[{\widehat T}_{\mu\nu} - {1\over{D-2}}\eta_{\mu\nu} {\widehat T}\right]\left[-\ln\left({\sqrt{p^2 + M^2}r_0\over 2}\right)-\gamma\right] -  \nonumber\\
 &&\,\,\,\, {M_P^{2-D}{\widehat T}\over a_{d-1}r_0^{d-2}}~ {{\eta_{\mu\nu}M^2 - p_\mu p_\nu}\over {(D-1)M^2}} \ln\left({\sqrt{p^2 + M^2}\over \sqrt{p^2 + M_h^2}}\right),\label{Hmunu-2}
\end{eqnarray}
where $\gamma$ is the Euler constant. Unlike the $d > 2$ case, in $d=2$ the term proportional to $(\eta_{\mu\nu}M^2 - p_\mu p_\nu)$ is no longer subleading. In fact, it becomes important when $M_h$ and $M$ are vastly different.

{}As in the codimension-1 case, let us write $H_{\mu\nu}$ as
\begin{equation}
 H_{\mu\nu} = a{\cal T}_{\mu\nu} + b{\cal T} + cp_\mu p_\nu{\cal T}~.
\end{equation}
Then we have
\begin{eqnarray}
 && a = {{\widehat M}_P^{2+d-D}\over {p^2 + \Omega_1(p)}}~,\\
 && b = -{{\widehat M}_P^{2+d-D}\over {D-d-1}}~{1 \over {p^2 + \Omega_1(p)}} - {{\widehat M}_P^{2+d-D}\over {(D-d-1)(D-d-2)}}~{1 \over {p^2 + \Omega_2(p)}}~,
\end{eqnarray}
where
\begin{eqnarray}
 &&\Omega_1(p)\equiv {a_{d-1}r_0^{d-2}M_P^{D-2}\over{\widehat M}_P^{D-d-2}}\left[-\ln\left({\sqrt{p^2+M^2}r_0\over 2}\right)-\gamma\right]^{-1},\\
 &&\Omega_2(p)\equiv {\Omega_1(p)(D-2)\over (d-1)(D-d-2)}\times\nonumber\\
 &&\,\,\,\,\,\,\,\times\left\{{(D-d-1)(D-2)\ln\left({\sqrt{p^2 + M^2}/\sqrt{p^2 + M_h^2}}\right)\over (D-1)(d-1)
 \left[-\ln\left({\sqrt{p^2 + M^2}r_0/2}\right)-\gamma\right]} - 1\right\}^{-1}.
\end{eqnarray}
Note that $\Omega_1(p) > 0$, so there is no tachyon pole in $a$. At $p^2\ll M^2$ we have
\begin{equation}
 \Omega_1(p)\approx m_c^2\equiv {a_{d-1}r_0^{d-2}M_P^{D-2}\over{\widehat M}_P^{D-d-2}}\left[-\ln\left({Mr_0\over 2}\right)-\gamma\right]^{-1}~,
\end{equation}
where $m_c\lsim H\ll M\sim M_P$. On the other hand, whether $\Omega_2(p)$ has a tachyon pole depends on the values of $M$ and $M_h$. First, note that $\Omega_1(p)\sim m_c^2$ for all momenta. If $M_h\gsim M$, then we have a tachyon pole at $p^2 = m_T^2$, where
\begin{equation}
 m_T^2\approx {m_c^2(D-2)\over (d-1)(D-d-2)}
 \left\{{(D-d-1)(D-2)\ln\left(M_h/M)\right)\over (D-1)(d-1)
 \left[-\ln\left({Mr_0/2}\right)-\gamma\right]} + 1\right\}^{-1}.
\end{equation}
Note that in this case $m_T^2\sim m_c^2$, and if we can ignore the tachyon for the reasons discussed above, then at momenta $p^2\gg m_c^2$ we have the $(D-d)$-dimensional tensor structure.

{}Next, let us assume that $M_h\ll M$. Furthermore, any tachyon pole would have to occur at $p^2\ll M^2$. Indeed, at $p^2\sim M^2$ we have $\Omega_2(p)\sim m_c^2$. Let us rewrite $\Omega_2(p)$ for momenta $p^2\ll M^2$ as follows:
\begin{equation}
 \Omega_2(p)\approx m_1^2 \left[\ln\left({m_0\over\sqrt{p^2 + M_h^2}}\right)\right]^{-1},
\end{equation}
where $m_0$ and $m_1$ are defined via
\begin{eqnarray}
 &&m_1^2\equiv {a_{d-1}r_0^{d-2}M_P^{D-2}\over{\widehat M}_P^{D-d-2}}~{{D-1}\over (D-d-1)(D-d-2)}~,\\
 &&\ln\left({M\over m_0}\right) = {(d-1)(D-1)\over(D-2)(D-d-1)}\left[-\ln\left({Mr_0\over 2}\right)-\gamma\right]~.\label{m0}
\end{eqnarray}
Note that, unlike the $d\geq 3$ cases, in $d=2$ we have only mild logarithmic hierarchy between the scales $m_1$ and $m_c$. Even for $r_0\sim 1/{\widehat M}_P$, this logarithmic hierarchy is less than 2 orders of magnitude. Therefore, the smallness of $m_c$ is not due to the smallness of $r_0$, but due to the smallness of $M_P/{\widehat M}_P$. Thus, in $D=6$ (a codimension-2 3-brane in 6D bulk) we have $m_1 = \sqrt{5\pi/6}M_P^2/{\widehat M}_P$, so $M_P$ must be roughly the inverse of 0.1 mm, as $m_c$ is smaller than $m_1$ only by an additional square-root-of-logarithmic factor, which is roughly 8 for $r_0\sim 1/{\widehat M}_P$ and even smaller when $r_0$ is larger. This is relevant because there is no reason to have a tiny $r_0$ as it does not help much in lowering $m_c$. In fact, {\em a priori} we could have $r_0$ as large as $1/M_P$. In this case we would have to work with full expressions as opposed to using the $r_0 M \ll 1$ approximation, which would make our computations more involved without necessarily changing any of the conclusions. Also, there would be no reason why we should not extend the Einstein-Hilbert term on the brane to the internal $S^1$. For the sake of simplicity we will therefore continue to assume that $r_0 M \ll 1$, but $r_0$ can be only an order of magnitude or two smaller than $1/M$, which is what is important in the following.

{}Thus, if $r_0$ is so small that $m_0\ll M$, then $m_0 \gg m_1$. This is because of the numeric factor on the r.h.s. of (\ref{m0}), which for $d=2$ and $D=6$ is 5/12, so even if $r_0\sim 1/{\widehat M}_P$, it only generates roughly 12-13 orders of magnitude hierarchy between $M$ and $m_0$. This implies that, if $m_0\ll M$, then for $M_h > m_0$ a tachyon pole in $1/(p^2 + \Omega_2(p))$ occurs at $p^2 = m_T^2$, where $m_T^2\approx m_1^2 / \ln(M_h/m_0)$. If $M_h \leq m_0$, then $\Omega_2(p)$ blows up at $p^2 = m_s^2$, where $m_s^2 \equiv m_0^2 - M_h^2$, and we still have a tachyon pole with $m_T^2\approx m_s^2/2 + \sqrt{m_s^2/4 + 2 m_0^2 m_1^2} \gg m_1$.

{}However, if $m_0$ is close to $M$ and $M_h\ll m_0$, then $\Omega_2(p)$ blows up at $p^2 =m_s^2\approx m_0^2$, {\em i.e.}, close to $M\sim M_P$ (and the would-be tachyon pole is just above $m_0$). {\em E.g.}, if $r_0 M = 1/10$, then $m_0\approx 0.4M$. What this really means is that our approximations (including the field theory treatment) break down above the scale $m_0$ and likely there is additional dynamics that must be taken into account, {\em e.g.}, stabilization of the size $r_0$ of the internal $S^1$ via some microscopic dynamics. However, there is no {\em infrared} tachyon in this case, which is always present in $d\geq 2$ when gravity in the bulk is massless \cite{Orientiworld}. So, at momenta $m_c^2\ll p^2\ll M^2$ we have $\Omega_2(p)\lsim m_1^2$ and the $(D-d)$-dimensional tensor structure as $b\approx -{\widehat M}_P^{3-D}/(D-d-2)p^2$. At $p^2 = m_s^2\approx m_0^2\sim M^2$ the tensor structure becomes $(D-d+1)$-dimensional -- as just mentioned, additional dynamics must be included above this scale. So, massive gravity in the bulk helps get rid of the infrared tachyon in the smoothed-out codimension-2 case by introducing additional terms into the graviton propagator that counterbalance the terms that give rise to the infrared tachyon in the massless case. Here we should mention that if $r_0\sim 1/M$, then this may also occur in $d\geq 3$ as the terms proportional to $(\eta_{\mu\nu}M^2 - p_\mu p_\nu)$ in (\ref{Hmunu-d}) are no longer subleading, albeit in this case there would be no reason not to extend the Einstein-Hilbert term to the internal $S^{d-2}$.

\subsection{Non-perturbative Dynamics, Nonzero Tension Branes, {\em etc.}}

{}The linearized analysis of the previous subsection assumed that the brane tension is 0. In codimension-1 there is no choice as there are no static solutions with positive brane tension. In codimension-2 we constructed positive-brane solutions in Sections 3 and 4 for the $\beta=1/2$ case. Because the background metric in this case factorizes as in (\ref{metric.D}), just as in the massless case discussed in \cite{Orientiworld}, the linearized analysis in this background is essentially the same as in the zero-tension case with all conclusions unchanged. In codimension-3 and higher nonzero positive brane tension solution should exist (for a continuous range of brane tension) once we include higher curvature terms, which complicate computations substantially and are outside of the scope of this paper. However, what we would like to comment on is non-perturbative dynamics.

{}As was discussed in \cite{Unitarity}, the non-perturbative Hamiltonian is bounded from below and the theory is unitary even for $\beta\not=1$, {\em i.e.}, when perturbatively the trace $h$ is a ghost. Simply put, the Hamiltonian is a non-polynomial function of the conjugate momentum, so for large values of the conjugate momentum -- precisely where the ``ghostliness" of $h$ becomes problematic perturbatively -- perturbative expansion is not valid to begin with. Does this mean that the above linearized analysis is invalid? The answer is that the linearized approximation is valid so long as the conjugate momentum for $h$ is small and $h$ itself is small -- more precisely, the same must be the case for all other components. For example, as we saw in the $\beta=1$ case, the trace $h$ is singular on the brane, and the perturbative approximation breaks down. Another issue is that, if we consider a point-like source on any codimension-$d$ brane, in the massive case the full non-perturbative solution is singular ({\em i.e.}, there is no ``horizon") unless we include higher-curvature terms. So, non-perturbative analysis of the metric of a point-like source on the brane requires inclusion of higher-curvature terms, unless such analysis is restricted to the asymptotic behavior far away from the source. In this regard, one might worry that, as the bulk gravity is massive, there might be present the vDVZ discontinuity and one might have to employ non-perturbative description even away from the source. However, as was argued in \cite{ZK-vDVZ}, the vDVZ discontinuity is an artifact attributable to the Fierz-Pauli mass term, {\em i.e.}, $\beta = 1$ (where the perturbative description breaks down as is), and for the other allowed values of $\beta$ perturbative asymptotic expansion is valid and there is no vDVZ discontinuity.

{}Finally, let us comment on ``self-accelerated" solutions. With massive gravity in the bulk, there are no such solutions. This is because the massive bulk gravity is a short-range force and asymptotically, far away from the brane, whatever is happening on the brane does not affect the bulk. Let us consider the metric of the form
\begin{equation}
 ds^2 = \exp(2A) ds_{D-d}^2 + \exp(2B) dr^2 + \exp(2C) r^2 \gamma_{ab}dx^a dx^b~,
\end{equation}
where $ds_{D-d}^2$ is the $(D-d)$-dimensional metric on the brane, $\gamma_{ab}$ is the metric on a unit sphere $S^{d-1}$, and
$A$, $B$ and $C$ are independent of the brane coordinates and depend only on the radial coordinate $r$ in the extra space. Because the bulk gravity is massive, far away from the brane $A$, $B$ and $C$ must go to 0, so asymptotically the space is $M_{D-d} \times R^d$, where $M_{D-d}$ is described by $ds_{D-d}^2$. So, $M_{D-d}$ can be flat as $R^{1,D-d-1} \times R^d$ is a solution to the bulk equations of motion. But $M_{D-d}$ cannot be (A)dS as (A)dS $\times R^d$ is not an asymptotic solution to the bulk equations of motion.

\section{Concluding Remarks}

{}Usually, massive gravity is thought of in the context of ``self-accelerated" solutions, where one tries to explain the apparent accelerated expansion of the Universe via an infrared modification of gravity -- in this case, small graviton mass. In this paper we propose and explore another application of massive gravity. We consider a Brane World scenario where we live on a 3-brane embedded in infinite-volume bulk with {\em massive} gravity propagating in the bulk. 

{}As in the case of massless gravity in the bulk, 4-dimensional Newton's law on the brane arises due to the presence of a 4D Einstein-Hilbert term in the brane world-volume action (which term can be induced via quantum corrections or be present at the classical level due to higher-curvature terms in the bulk). Not only does the 4D Einstein-Hilbert term on the brane shield the brane matter from strong bulk gravity (which is the case whether bulk gravity is massive or massless), but also from large bulk graviton mass, which is one of the observations of this paper. Because of this shielding, the bulk graviton can be
much heavier than the inverse Hubble size (which is roughly the upper bound on 4-dimensional graviton mass). In fact, the bulk graviton can be as heavy as the bulk Planck scale $M_P$, whose lower bound is roughly the inverse of 0.1 mm. Furthermore, gravity on the brane does not become higher-dimensional at large
distances. Instead, at distance scales above the bulk Planck length, gravity on the
brane behaves as 4D gravity with small graviton mass roughly of order or below
the inverse Hubble size. The aforementioned features are universal when we have massive gravity in the bulk, irrespective of the number of extra dimensions, or the codimension of the brane.

{}There are, however, codimension-dependent features as well. Thus, in the case of massless gravity in the bulk, in the codimension-1 case (that is, in the case of a 3-brane embedded in 5D bulk with the infinite fifth dimension), while Newton's law of gravitation is recovered on the brane due to the presence of the 4D Einstein-Hilbert term in the brane world-volume action, the tensor structure of the graviton propagator is always 5-dimensional. However, unlike the massless case, with massive gravity in the bulk we can have 4D tensor structure on a codimension-1 brane. This is achieved by considering graviton mass term in the bulk $-(M^2/4)\left[h^{MN}h_{MN} - \beta h^2\right]$ with $\beta\neq 1$. In fact, the trace $h=h^M_M$ does not decouple from the brane matter sources even at $\beta = 1$ and, in fact, the perturbative expansion breaks down altogether in the case of the Fierz-Pauli mass term ($\beta=1$). (And this is the case irrespective of the codimension.) For $\beta < 1$, perturbatively the trace $h$ is a propagating ghost. However, non-perturbatively the Hamiltonian is bounded from below and there is no ghost. This is how massive gravity in the bulk circumvents the perturbative argument that to obtain 4D tensor structure on a 3-brane embedded in infinite-volume 5D bulk some ``unconventional" states are needed -- the trace $h$ is precisely this ``unconventional" state by the virtue of its naive perturbative ``ghostliness", while non-perturbatively the theory is unitary. Another remark about the codimension-1 case is that, with massless gravity in the bulk, above the cross-over scale $r_c\sim M_P^3 / {\widehat M}_P^2$ the gravity on the brane becomes 5-dimensional (here ${\widehat M}_P$ is the 4D Planck mass). With massive gravity in the bulk, there is no cross-over to 5D gravity; instead, the graviton propagator on the brane is modified by a small mass of order $m_c\sim \sqrt{M_P^3 M} / {\widehat M}_P$, so we have $m_c\sim M_P^2/{\widehat M}_P$ when the bulk graviton mass $M \sim M_P$, and the lower bound on the bulk Planck mass $M_P$ is roughly the inverse of 0.1 mm, which is how heavy the bulk graviton {\em a priori} can be.

{}In higher codimension $d >1$ cases, due to the fact that the spherically symmetric Green's function in $d$ dimensions is singular at the origin for $d>1$, the brane must be partially smoothed out, so the brane world-volume becomes $R^{1,3} \times S^{d-1}$, where $S^{d-1}$ has small radius $r_0$. Because of this, in the case of massless gravity in the bulk there is always an infrared tachyon with negative mass-squared of order $m_T^2\sim m_c^2$, where $m_c^2\sim r_0^{d-2}M_P^{d+2}/{\widehat M}_P^2$. (Here $r_0^{d-2}$ is replaced by $1/\log(R/r_0)$ for $d=2$, where $R$ is an infrared cut-off scale.) However, massive gravity in the bulk helps get
rid of the infrared tachyon by introducing additional terms into the graviton propagator that counterbalance the terms that
give rise to the infrared tachyon in the massless case. (We discuss this in detail in the codimension-2 case and comment on the $d > 2$ cases.) Also, in codimension-2 cases there exist nonsingular solutions with a flat 3-brane for a continuous positive range of the brane tension.

{}Finally, let us mention that with massive gravity in the bulk the
effects of the brane dynamics on the bulk are exponentially suppressed away
from the brane. One consequence of this is that there are no ``self-accelerated" solutions.

\subsection*{Acknowledgments}
{}I would like to thank Olindo Corradini and Alberto Iglesias for reading an early version of the manuscript and valuable comments.


\end{document}